\documentclass[a4paper,12pt]{article}

\usepackage{graphics,psfrag}
\usepackage{amsmath,amsthm,amssymb,epsfig,euscript,array,cite,cancel,color}
\usepackage{cases,empheq}
\usepackage[colorlinks=true]{hyperref}     
\usepackage{verbatim}
\usepackage{mciteplus}
\usepackage{soul}   

\usepackage[usenames,dvipsnames,svgnames,table]{xcolor}

\if{}
\usepackage[paper=a4paper,dvips,top=2.54cm,left=2.74cm,right=2.34cm,
    foot=1cm,bottom=2.54cm]{geometry}
\fi

\theoremstyle{definition}

\theoremstyle{remark}

\newcounter{multieqs}

\newcommand{\be}{\begin{equation}}
\newcommand{\ee}{\end{equation}}
\newcommand{\eq}[1]{(\ref{#1})}
\newcommand{\bit}{\begin{itemize}}  \newcommand{\eit}{\end{itemize}}

\newcommand{\bra}[1]{\langle #1|}
\newcommand{\ket}[1]{|#1 \rangle}
\newcommand{\ipr}[2]{\langle #1 | #2 \rangle}

\newcommand{\bm}[1]{\mbox{\boldmath $#1$}}
\newcommand{\rf}[1]{(\ref{#1})}

\def\bd{\begin{document}}
\def\ed{\end{document}}
\def\nn{\nonumber}
\def\bea{\begin{eqnarray}}
\def\eea{\end{eqnarray}}
\let\bm=\bibitem

\def\la{\langle}
\def\ra{\rangle}

\def\npb#1#2#3{Nucl. Phys. {\bf{B#1}} #3 (#2)}
\def\plb#1#2#3{Phys. Lett. {\bf{#1B}} #3 (#2)}
\def\prl#1#2#3{Phys. Rev. Lett. {\bf{#1}} #3 (#2)}
\def\prd#1#2#3{Phys. Rev. {D \bf{#1}} #3 (#2)}
\def\cmp#1#2#3{Comm. Math. Phys. {\bf{#1}} #3 (#2)}
\def\cqg#1#2#3{Class. Quantum Grav. {\bf{#1}} #3 (#2)}
\def\nppsa#1#2#3{Nucl. Phys. B (Proc. Suppl.) {\bf{#1A}}#3 (#2)}
\def\ap#1#2#3{Ann. of Phys. {\bf{#1}} #3 (#2)}
\def\ijmp#1#2#3{Int. J. Mod. Phys. {\bf{A#1}} #3 (#2)}
\def\rmp#1#2#3{Rev. Mod. Phys. {\bf{#1}} #3 (#2)}
\def\mpla#1#2#3{Mod. Phys. Lett. {\bf A#1} #3 (#2)}
\def\jhep#1#2#3{J. High Energy Phys. {\bf #1} #3 (#2)}
\def\atmp#1#2#3{Adv. Theor. Math. Phys. {\bf #1} #3 (#2)}

\def\N{{\cal N}}
\def\sst{\scriptscriptstyle}
\def\thetabar{\bar\theta}
\def\Tr{{\rm Tr}}
\def\one{\mbox{1 \kern-.59em {\rm l}}}

\def\a{\alpha}      \def\da{{\dot\alpha}}  \def\dA{{\dot A}}
\def\b{\beta}       \def\db{{\dot\beta}}
\def\g{\gamma}  \def\G{\Gamma}  \def\dc{{\dot\gamma}}
\def\d{\delta}  \def\D{\Delta}  \def\ddt{\dot\delta}
\def\e{\epsilon}        \def\ve{\varepsilon}
\def\f{\phi}    \def\F{\Phi}    \def\vvf{\f}
\def\h{\eta}
\def\k{\kappa}
\def\l{\lambda} \def\L{\Lambda}
\def\m{\mu} \def\n{\nu}
\def\o{\omega}
\def\p{\pi} \def\P{\Pi}
\def\r{\rho}
\def\s{\sigma}  \def\S{\Sigma}
\def\t{\tau}
\def\th{\theta} \def\Th{\Theta} \def\vth{\vartheta}
\def\X{\Xeta}
\def\z{\zeta}

\def\na{\nabla}

\def\cA{{\cal A}} \def\cB{{\cal B}} \def\cC{{\cal C}}
\def\cD{{\cal D}} \def\cE{{\cal E}} \def\cF{{\cal F}}
\def\cG{{\cal G}} \def\cH{{\cal H}} \def\cI{{\cal I}}
\def\cJ{{\cal J}} \def\cK{{\cal K}} \def\cL{{\cal L}}
\def\cM{{\cal M}} \def\cN{{\cal N}} \def\cO{{\cal O}}
\def\cP{{\cal P}} \def\cQ{{\cal Q}} \def\cR{{\cal R}}
\def\cS{{\cal S}} \def\cT{{\cal T}} \def\cU{{\cal U}}
\def\cV{{\cal V}} \def\cW{{\cal W}} \def\cX{{\cal X}}
\def\cY{{\cal Y}} \def\cZ{{\cal Z}}

\def\ua{\underline{\alpha}}
\def\uc{\underline{\phantom{\alpha}}\!\!\!\gamma}
\def\um{\underline{\mu}}
\def\ud{\underline\delta}
\def\ue{\underline\epsilon}
\def\una{\underline a}\def\unA{\underline A}
\def\unb{\underline b}\def\unB{\underline B}
\def\unc{\underline c}\def\unC{\underline C}
\def\und{\underline d}\def\unD{\underline D}
\def\une{\underline e}\def\unE{\underline E}
\def\unf{\underline{\phantom{e}}\!\!\!\! f}\def\unF{\underline F}
\def\unm{\underline m}\def\unM{\underline M}
\def\unn{\underline n}\def\unN{\underline N}
\def\unp{\underline{\phantom{a}}\!\!\! p}\def\unP{\underline P}
\def\unq{\underline{\phantom{a}}\!\!\! q}
\def\unQ{\underline{\phantom{A}}\!\!\!\! Q}
\def\unH{\underline{H}}

\def\As {{A \hspace{-6.4pt} \slash}\;}
\def\bs {{b \hspace{-6.4pt} \slash}\;}
\def\Ds {{D \hspace{-6.4pt} \slash}\;}
\def\Gts {{\Gt \hspace{-6.4pt} \slash}\;}
\def\ds {{\del \hspace{-6.4pt} \slash}\;}
\def\ss {{\s \hspace{-6.4pt} \slash}\;}
\def\ks {{ k \hspace{-6.4pt} \slash}\;}
\def\ps {{p \hspace{-6.4pt} \slash}\;}
\def\xs {{x \hspace{-6.4pt} \slash}\;}
\def\pas {{{p_1} \hspace{-6.4pt} \slash}\;}
\def\pbs {{{p_2} \hspace{-6.4pt} \slash}\;}
\def\cFs {{{\cal F} \hspace{-6.4pt} \slash}\;}

\def\Ah{{\hat{A}}}
\def\Dh{{\hat{D}}}
\def\Gh{{\hat{G}}}
\def\Fh{{\hat{F}}}
\def\Ih{{\hat{I}}}
\def\Jh{{\hat{J}}}
\def\Kh{{\hat{K}}}
\def\Lh{{\hat{L}}}
\def\Ph{{\hat{P}}}
\def\Rh{{\hat{R}}}
\def\Vh{{\hat{V}}}
\def\Xh{{\hat{X}}}

\def\ah{{\hat{a}}}
\def\bh{{\hat{b}}}
\def\ch{{\hat{c}}}
\def\gh{{\hat{g}}}
\def\dh{{\hat{d}}}
\def\hh{{\hat{h}}}
\def\uh{{\hat{u}}}
\def\vh{{\hat{v}}}
\def\xh{{\hat{x}}}
\def\yh{{\hat{y}}}
\def\zh{{\hat{z}}}
\def\ph{{\hat{p}}}
\def\thh{{\hat{t}}}
\def\xih{\hat{\xi}}
\def\Psih{\hat{\Psi}}
\def\mh{{\hat{m}}}
\def\nh{{\hat{n}}}
\def\ih{{\hat{i}}}
\def\jh{{\hat{j}}}
\def\kh{{\hat{k}}}
\def\aah{{\hat{\alpha}}}
\def\bbh{{\hat{\beta}}}
\def\ggh{{\hat{\gamma}}}
\def\llh{{\hat{\ell}}}
\def\ph{{\hat{p}}}

\def\psit{\tilde{\psi}}
\def\Psit{\tilde{\Psi}}
\def\Psibt{\tilde{\bar{Psi}}}

\def\delt{\tilde{\delta}}
\def\Phit{\tilde{\Phi}}
\def\Phitb{\overline{\tilde{Phi}}}
\def\tht{\tilde{\th}}
\def\lt{\tilde{\l}}
\def\chit{\tilde{\chi}}
\def\phit{\tilde{\phi}}

\def\At{\tilde{A}}
\def\Bt{\tilde{B}}
\def\Ct{\tilde{C}}
\def\Dt{\tilde{D}}
\def\Et{\tilde{E}}
\def\Ft{\tilde{F}}
\def\Gt{\tilde{G}}
\def\Ht{\tilde{H}}
\def\It{\tilde{I}}
\def\Jt{\tilde{J}}
\def\Qt{\tilde{Q}}
\def\Rt{\tilde{R}}
\def\Mt{\tilde{M }}
\def\Nt{\tilde{N}}
\def\St{\tilde{S}}
\def\Vt{\tilde{V}}
\def\Xt{\tilde{X}}
\def\at{\tilde{a}}
\def\ct{\tilde{c}}
\def\dt{\tilde{d}}
\def\htt{\tilde{h}}
\def\ft{\tilde{f}}
\def\gt{\tilde{g}}
\def\pt{\tilde{p}}
\def\qt{\tilde{q}}
\def\vt{\tilde{v}}
\def\nt{\tilde{n}}
\def\ut{\tilde{u}}
\def\wt{\tilde{w}}
\def\zt{\tilde{z}}
\def\xt{\tilde{x}}
\def\yt{\tilde{y}}
\def\Psit{\tilde{\Psi}}
\def\vphit{\tilde{\varphi}}

\def\eb{\bar{\epsilon}}
\def\delb{\bar{\partial}}
\def\thb{\bar{\theta}}
\def\mub{\bar{\mu}}
\def\lamb{\bar{\l}}
\def\psib{\bar{\psi}}
\def\sb{\bar{\sigma}}
\def\xib{\bar{\xi}}
\def\chib{\bar{\chi}}

\def\Psib{\bar{\Psi}}
\def\Phib{\bar{\Phi}}
\def\Lamb{\bar{\Lambda}}
\def\Sb{{\overline \Sigma}}
\def\cb{\bar{c}}
\def\hb{\bar{h}}
\def\qb{\bar{q}}
\def\wb{\bar{w}}
\def\ub{\bar{u}}
\def\zb{{\bar{z}}}
\def\Hb{\bar{H}}
\def\Qb{{\bar Q}}
\def\Omegab{\overline{\Omega}}
\def\ob{\overline{\omega}}

\def\Ab{{\overline A}} \def\Bb{{\overline B}} \def\Cb{{\overline C}}
\def\Db{{\overline D}} \def\Eb{{\overline E}} \def\Fb{{\overline F}}
\def\Gb{{\overline G}}
\def\Ib{{\overline I}}
\def\Jb{{\overline J}} \def\Kb{{\overline K}} \def\Lb{{\overline L}}
\def\Mb{{\overline M}} \def\Nb{{\overline N}} \def\Ob{{\overline O}}
\def\Pb{{\overline P}}  \def\Rb{{\overline R}}
 \def\Tb{{\overline T}} \def\Ub{{\overline U}}
\def\Vb{{\overline V}} \def\Wb{{\overline W}} \def\Xb{{\overline X}}
\def\Yb{{\overline Y}} \def\Zb{{\overline Z}}

\def\fb{{\overline f}}
\def\gb{{\overline g}}
\def\mb{{\overline m}}
\def\lb{{\overline l}}
\def\yb{{\overline y}}

\def\ldel{{\overleftarrow{\del}}}
\def\rdel{{\overrightarrow{\del}}}
\def\ldeldel{{\overleftarrow{\del^2}}}
\def\rdeldel{{\overrightarrow{\del^2}}}
\def\ldelb{{\overleftarrow{\bar{\del}}}}
\def\rdelb{{\overrightarrow{\bar{\del}}}}

\def\ba{{\bf a}}
\def\bk{{\bf k}}
\def\bl{{\bf l}}
\def\bp{{\bf p}}
\def\bq{{\bf q}}
\def\br{{\bf r}}
\def\bt{{\bf t}}
\def\bu{{\bf u}}
\def\bv{{\bf v}}
\def\bx{{\bf x}}
\def\by{{\bf y}}
\def\bR{{\bf R}}
\def\bV{{\bf V}}

\def\bone{{\bf 1}}

\def\va{{\vec a}}
\def\vk{{\vec k}}
\def\vp{{\vec p}}
\def\vq{{\vec q}}
\def\vx{{\vec x}}
\def\vy{{\vec y}}
\def\vu{{\vec u}}
\def\vv{{\vec v}}

\def\vs{{\vec \sigma}}
\def\vtau{{\vec \tau}}

\newcommand{\ov}[1]{\overrightarrow{#1}}

\def\frA{\mathfrak{A}}
\def\frB{\mathfrak{B}}
\def\frC{\mathfrak{C}}
\def\frD{\mathfrak{D}}
\def\frE{\mathfrak{E}}
\def\frF{\mathfrak{F}}
\def\frG{\mathfrak{G}}
\def\frH{\mathfrak{H}}
\def\frM{\mathfrak{M}}
\def\frN{\mathfrak{N}}
\def\frR{\mathfrak{R}}
\def\frW{\mathfrak{W}}

\def\fra{\mathfrak{a}}
\def\frb{\mathfrak{b}}
\def\frf{\mathfrak{f}}
\def\frg{\mathfrak{g}}
\def\frh{\mathfrak{h}}
\def\frl{\mathfrak{l}}
\def\frs{\mathfrak{s}}
\def\fri{\mathfrak{i}}
\def\frj{\mathfrak{j}}

\def\ma{\mathfrak{a}}
\def\mg{\mathfrak{g}}
\def\mh{\mathfrak{h}}
\def\mR{\mathfrak{R}}
\def\mN{\mathfrak{N}}

\def\d{\delta}\def\D{\Delta}\def\ddt{\dot\delta}

\def\pa{\partial} \def\del{\partial}
\def\xx{\times}
\def\uno{\mbox{1 \kern-.59em {\rm l}}}

\def\trp{^{\top}}
\def\inv{^{-1}}
\def\dag{{^{\dagger}}}
\def\pr{^{\prime}}

\def\rar{\rightarrow}
\def\lar{\leftarrow}
\def\lrar{\leftrightarrow}

\newcommand{\0}{\,\!}      
\def\one{1\!\!1\,\,}
\def\im{\imath}
\def\jm{\jmath}

\newcommand{\tr}{\mbox{tr}}
\newcommand{\slsh}[1]{/ \!\!\!\! #1}

\def\vac{|0\rangle}
\def\lvac{\langle 0|}

\def\hlf{\frac{1}{2}}
\def\ove#1{\frac{1}{#1}}

\def\Box{\square}
\def\CC {\mathbb{C}}
\def\FF {\mathbb{F}}
\def\RR{\mathbb{R}}
\def\NN{\mathbb{N}}
\def\ZZ{\mathbb{Z}}
\def\bb#1{{\bf #1}}
\def\bcomment#1{}
\def\bfhat#1{{\bf \hat{#1}}}
\def\VEV#1{\left\langle #1\right\rangle}

\newcommand{\ex}[1]{{\rm e}^{#1}} \def\ii{{\rm i}}

\newcommand{\lrbrk}[1]{\left(#1\right)}
\newcommand{\lrsbrk}[1]{\left[#1\right]}
\newcommand{\sfrac}[2]{{\textstyle\frac{#1}{#2}}}

\def\stw{{\sqrt{2}}}

\def\rf {{\rm f}}
\def\ri {{\rm i}}
\def\rj {{\rm j}}
\def\rk {{\rm k}}
\def\rl {{\rm l}}
\def\rs {{\scriptscriptstyle \rm S}}
\def\rt {{\scriptscriptstyle \rm T}}

\def\rQ {{\scriptscriptstyle \rm \cQ}}
\def\rR {{\scriptscriptstyle \rm \cR}}

\def\cQb{{\cal \Qb}}
\def\cRb{{\cal \Rb}}
\def\cWb{{\cal \Wb}}

\def\fd {{\rm N}}
\def\afd {{\overline{\rm N}}}

\def \II {I\hspace{-.1em}I\hspace{.1em}}
\def \IIA {\mbox{\II A\hspace{.2em}}}
\def \IIB {\mbox{\II B\hspace{.2em}}}
\def \gs {g^s}
\def \ls {\lambda^s}

\def \I {{\cal I}}
\def \qs {q\hspace{-.53em}/\hspace{.15em}}
\def \ks {k\hspace{-.53em}/\hspace{.15em}}
\def \YM {{\mbox{\tiny YM}}}
\def \gym {g_{\YM}}

\def \Lc {\L_c}
\def\IR{\relax{\rm I\kern-.18em R}}
\def \id {{\bf 1}}

\def\cci{\ell}
\def\ccj{\ell'}

\def \thbb{\overline{\th\th}}
\newcommand \ol{\overline}
\def \lamb{\bar{\lambda}}
\def \vphi{\varphi}
\def \lambh{\hat{\bar{\lambda}}}
\def \lh{\hat{\lambda}}
\def \dd{\ddagger}

\def \ad {\dot{a}}
\def \bd {\dot{b}}
\def \cd {\dot{c}}
\def  \ddd {\dot{d}}
\def \ed {\dot{e}}
\def \fd {\dot{f}}
\def \Bh {\hat{B}}
\def \zm {{(0)}}
\def \nz {{(\text{KK})}}
\def \3{{(3)}}
\def \diag {\text{diag}}
\def \inm {{(m^{-1})}}

\def\eh{{\hat{e}}}
\def\fh{{\hat{f}}}
\renewcommand{\mh}{\hat{m}}
\def\theequation{\thesection.\arabic{equation}}

\def\adj{\text{adj}}
\def\co{\text{co}}
\def\6{{\text{(6)}}}
\def\5{{\text{(5)}}}

\author{Sheng-Lan Ko
\footnote{ko.shenglan@gmail.com}$~^{a,b}$
\,and Pichet Vanichchapongjaroen
\footnote{pichetv@nu.ac.th}$~^b$
\\
\\
{\small $^a$ \it Centre for Particle Theory
and Department of Mathematical Sciences,}
\\
{\small \it Durham University, Durham, DH1 3LE, UK   }
\\
\\
{\small $^b$ \it The Institute for Fundamental Study ``The Tah Poe Academia Institute",}
\\
{\small\it Naresuan University, Phitsanulok 65000, Thailand}
}

\title{\bf Towards 2+4 formulation of M5-brane}

\date{}
\begin{document}
\maketitle

\abstract{
We present {\color{black}an} attempt to formulate an action
for the worldvolume theory of a single M5-brane, based on the splitting of the six worldvolume directions into 2+4, which breaks manifest Lorentz invariance from $SO(1,5)$ to $SO(1,1)\times SO(4)$.
To this end, an action for the free six--dimensional (2,0) chiral tensor multiplet,
and separately, a nonlinearly interacting chiral 2-form action are constructed.
By studying the Lagrangian formulation for the chiral 2-form with 2+4 splitting,
it is suggested that, if exists, the modified diffeomorphism of the theory
on curved six--dimensional space--time is less trivial
than its 1+5 and 3+3 counterpart, thus hindering the coupling of the chiral 2-form to the induced metric on the worldvolume of the M5-brane. 
We discuss difficulties of further generalisation of the theory. 
Finally, in terms of Hamiltonian analysis, we show that the naively gauge-fixed failed-PST-covariantised Lagrangian has the correct number of degrees of freedom, 
and satisfies the hyper--surface deformation algebra.
}
\thispagestyle{empty}
\newpage
\tableofcontents

\setcounter{equation}0
\section{Introduction}
Recently, an alternative M5-brane action in a generic eleven-dimensional supergravity background was constructed in \cite{Ko:2013dka}
with the aim of better understanding the connection of the original M5-brane action \cite{Bandos:1997ui,Aganagic:1997zq}
to the 5-brane proposal of \cite{Ho:2008nn, Ho:2008ve} based on the three-dimensional
Bagger-Lambert-Gustavsson model \cite{Bagger:2006sk, Bagger:2007jr, Gustavsson:2007vu}
with the gauge symmetry of a 3d volume preserving diffeomorphism.
In \cite{Ko:2013dka} it was shown that the field equations derived from the new action are equivalent
to the ones deduced from the superembedding approach  \cite{Howe:1996yn, Howe:1997fb} and hence to the equations of motion which follow from the original action \cite{Bandos:1997gm}.

The difference between the two M5-brane actions is that in the original action of \cite{Bandos:1997ui,Aganagic:1997zq} the 6-dimensional M5-brane worldvolume gets split into 1+5 directions and the manifest 6d space-time invariance is maintained by the presence of a single auxiliary scalar field, while in the action of \cite{Ko:2013dka} the 6d worldvolume is effectively split into 3+3 directions and the manifest 6d space-time invariance is maintained by the introduction of a triplet of auxiliary scalar fields \cite{Pasti:2009xc}.

Different formulations of the theory may allow one to gain different insights into its structure. The action of \cite{Ko:2013dka}, for instance, in addition to its relation to the BLG model, can also be useful for studying  M2-M5 bound states discussed e.g. in \cite{Niarchos:2014maa,Mori:2014tca}.

The Lagrangian formulation of self-dual or duality-symmetric fields is essentially not unique,
but is related to different possible ways of tackling the issue of (non-manifest) space-time invariance of the duality-symmetric actions (see e.g.
\cite{Zwanziger:1970hk,Deser:1976iy,Henneaux:1987hz,Henneaux:1988gg,Schwarz:1993vs,Perry:1996mk,Maznytsia:1998xw,Belov:2006jd,Ho:2008nn}).
Various possible ways of constructing actions which produce the (self)-duality relations as (a consequence of) equations of motion
by effectively splitting
$d$-dimensional space-time into $p$- and $q$-dimensional subspaces, with $d=p+q$, were explored for free theories {\color{black} in flat space} in \cite{Chen:2010jgb,Huang:2011np}. In these formulations only $SO(1,p-1)\times SO(q)$ subgroup of the $SO(1,d-1)$ Lorentz symmetry is manifest, while the complete 6d invariance is realized in a non--manifest (modified) form.
Recently, an action for IIB D=10 supergravity, containing a chiral four--form  gauge field $A_4$ whose field strength $F_5=dA_4$ is self--dual, was constructed in a 5+5 split formulation which originated from an $E_{6(6)}$ Exceptional Field Theory \cite{Hohm:2013vpa}. This formulation is alternative to the earlier constructed D=1+9 IIB supergravity action \cite{Dall'Agata:1997ju,Dall'Agata:1998va}.

The actions with different space-time splitting are generically inequivalent off-shell,
as was shown for the $6=1+5$ and $6=3+3$ cases in \cite{Pasti:2009xc,Ko:2013dka}.
Different off-shell inequivalent formulations may be useful for studying the dynamics of duality-symmetric fields in topologically non-trivial backgrounds \cite{Bekaert:1998yp,Belov:2006jd,Bandos:2014bva,Isono:2014bsa} and their
quantization \cite{Witten:1996hc,Dolan:1998qk,Henningson:1999dm,Witten:1999vg,Bekaert:2000rh,Belov:2006jd,Chen:2013gca}.
{\color{black}Potentially, these $6=p+q$ chiral 2-form theories in six dimensional
Minkowski space may be extended to describe
the worldvolume theory of the M-theory five brane, as it was carried out in
\cite{Bandos:1997ui,Aganagic:1997zq} and \cite{Ko:2013dka}
for the cases of 6=1+5 and 6=3+3. }

The above reasoning has motivated us to complete the list of different Lagrangian formulations of the M5-brane by constructing its action with an effective 2+4 splitting of the 6d worldvolume. Another motivation is that this form of the action for the Abelian $\mathcal N=(2,0)$ $d=6$ theory would provide us with an appropriate off--shell starting point for its topological twisting considered recently in \cite{Anderson:2013hpa,Gran:2014lia}.

To construct the `2+4' M5-brane action one may try to follow the same strategy as that for the `3+3' action \cite{Ko:2013dka}:
first deform the action \cite{Chen:2010jgb} for the free chiral two--form  to a nonlinear one, couple it to 6d gravity,
embed the M$5$-brane worldvolume into $D=11$ supergravity background and finally search for the kappa-symmetry invariant form of the non-linear action.

It turns out, however, that these steps cannot be accomplished in full.
Although it is possible to extend the free 2+4 action of \cite{Chen:2010jgb}
by supersymmetrising it in the worldvolume, or separately by making it non-linear,
there are obstacles in carrying out further steps.
Most notably, it is not clear how to covariantise the 2+4 action.
Nevertheless, as suggested by Hamiltonian analysis, coupling to 6d gravity might be possible.

In comparison with its previous counterparts, the `2+4' self--dual Lagrangian formulation for the chiral 2--form field has several new features and complications.
Namely, some of the gauge symmetries of the action become semi-local\footnote{Semi-local symmetries have previously appeared also in other formulations of duality-symmetric fields in different dimensions (see e.g. \cite{Cherkis:1997bx,Maznytsia:1998xw,Sevrin:2013nca}) and topologically non-trivial backgrounds \cite{Bekaert:1998yp,Bandos:2014bva,Isono:2014bsa}.}. For these semi-local transformations to be \emph{gauge} symmetries, the time direction of the $d=2+4$ worldvolume should be in the two-dimensional subspace, thus breaking 6d
`space-time democracy', though the action does possess a (modified)
6d Lorentz invariance.
The structure of the M5--brane action with `2+4' splitting (if found) is
anticipated to be
much more complicated in comparison with a Born-Infeld-like structures of the actions of \cite{Bandos:1997ui,Aganagic:1997zq} and \cite{Ko:2013dka}. A defining function of components of the chiral tensor field strength which enters the action should satisfy an algebraic equation of the sextic order which can only be solved perturbatively.

The problem of covariantising the 2+4 split action for the chiral 2-form may be related to issues with topological twisting of the Abelian 6d, $\mathcal N=(2,0)$ theory considered in \cite{Anderson:2013hpa,Gran:2014lia}.

The paper is organized as follows. In Section \ref{sec:free}, we review the free non-covariant `2+4' chiral 2-form gauge field Lagrangian,
and extend it to describe an Abelian $\mathcal N=(2,0)$ $d=6$ chiral supermultiplet.
The derivation of a new nonlinear action for the $d=6$ chiral 2--form field is considered in Section \ref{sec:nonlinear}.
In Section \ref{sec:iss}, we discuss obstacles to get full M5-brane action with `2+4' splitting,
as well as discussing a possible way out encouraged by Hamiltonian analysis.
In Conclusion  we summarize the results and
discuss open issues, and possible future directions.
In Appendix \ref{app:SDSE}, we give the detailed proof of the equivalence
of the self-duality equations derived from the new nonlinear `2+4' action
with the ones in the superembedding approach.
In Appendix \ref{app:SC} we explicitly check that a nonlinear function of the components of the chiral field strength in the `2+4' action
satisfies the constraint required by the non--manifest 6d
Lorentz invariance of the action.

\setcounter{equation}0
\subsubsection*{Basic notation and conventions}
The 6d  Minkowski metric has the almost plus signature, $x^\mu$ ($\mu=0,1,\cdots, 5$) stand for the 6d space-time coordinates. The chiral gauge field is denoted by $B_2(x)=\frac 12 dx^\mu dx^\nu B_{\nu\mu}(x)$.
We use the convention that the functional derivative and the variational derivative 
are related by,
\be
\d F_{\m_1\cdots\m_p} = \d F_{\n_1\cdots\n_p} \, \ove{p!}\, \frac{\pa F_{\m_1\cdots\m_p}}{\pa F_{\n_1\cdots\n_p}}  
= \d F_{\n_1\cdots\n_p} \,  \frac{\d F_{\m_1\cdots\m_p}}{\d F_{\n_1\cdots\n_p}}, 
\ee
for the variation of a $p-$form $F_{\m_1\cdots\m_p}$.

\setcounter{equation}0
\section{Free chiral 2--form theory with non-manifest $6d$ Lorentz-invariance} \label{sec:free}
We will now review the non-manifestly 6d Lorentz invariant quadratic chiral
2-form action in {\color{black} six dimensional}
Minkowski space and then extend it to an action describing the $\mathcal N=(2,0)$ tensor supermultiplet with five scalars and a sixteen--component fermion.

\subsection{Free theory}
We are interested in the derivation of the self-duality condition
\be\label{H=*H}
H_{\mu\nu\rho}=\frac 16 \varepsilon_{\mu\nu\rho\lambda_1\lambda_2\lambda_3}\,H^{\lambda_1\lambda_2\lambda_3}=\tilde H_{\mu\nu\rho}
\ee
on the field strength $H_3=dB_2$
from a 6d Lagrangian with a 2+4 splitting of six-dimensional tensor indices \cite{Chen:2010jgb}.
Here $\e^{012345}=-\e_{012345}=1$.

Let us perform the following 2+4 splitting of $H_{\mu\nu\rho}$
\be\label{H3(2+4)}
H_{\mu\nu\rho}=(H_{abj},H_{ijk}, H_{aij}),\qquad  a,b,c,\cdots = 0,5; \qquad i,j,k,\cdots = 1,2,3,4.
\ee
Then, the Hodge-dual field-strength $\tilde H_{\mu\nu\rho}$ splits as follows
\be
\varepsilon_{\mu_1\cdots\mu_6}\quad \Rightarrow\quad \e_{ab ijkl} = \e_{ab}\e_{ijkl},
\ee
\be \label{flatDual}
\Ht_{ab i} = \ove{3!} \e_{ab}\varepsilon_{ijkl}H^{jkl}, \quad
\Ht_{a ij} = \hlf \e_{ab}\varepsilon_{ijkl }H^{bkl}, \quad
\Ht_{ijk} = \hlf\e_{ijkl}\varepsilon_{ab}H^{ab l}.
\ee
The quadratic action which produces \eqref{H=*H} has the following form \cite{Chen:2010jgb}
\be\label{quadraticS}
S =
	-\int d^6x\lrbrk{\hlf\Ht_{ab i}H^{ab i} + \ove{4}H_{a ij}H^{a ij} + \ove{6}H_{ijk}H^{ijk}}.
\ee
The action has the local gauge symmetry
\be \label{Lsym1}
\d B_{ab} = \Omega_{ab}(x^\mu),
\ee
where $\Omega_{ab}(x^\mu)$ are arbitrary functions, which suggests that the $B_{ab}$ components of $B_2$ are Stueckelberg--like fields (they enter the above action only under a total derivative).

In addition, as we have found, the action is also invariant under the following
semi-local transformations
\be \label{LSMsym}
\d B_{ai} = \F_{a i}(x^b,x^j)
\ee
whose parameters $\F_{a i}$ are restricted to satisfy the anti-self-duality condition
\be\label{paraCond}
\pa^{[i}\F^{k] a}= - \ove{2}\e^{a b}\e^{ikjl} \del_j \F_{b l},
\quad\text{so that}\quad \pa_k \pa^{[k} \F^{i] a} = 0,
\ee
\emph{i.e.} $\F_{a i}$ obey the differential equation in the four-dimensional subspace parametrized by the coordinates $x^i$.

We should check that, though being semi-local, the transformations \eqref{LSMsym} form a genuine gauge symmetry which will allow us to get rid of redundant degrees of freedom\footnote{The presence of this semi-local gauge symmetry is effectively translated into the choice of appropriate boundary conditions for integration functions considered in \cite{Chen:2010jgb}.}.

A semi-local symmetry is a fully-fledged {\textit{gauge symmetry}}
if its associated Noether charge vanishes (at least) on the mass shell \cite{Henneaux:1992ig}.
The conserved Noether current associated with \eqref{LSMsym} is
\be \label{Noecurr}
j^\m = \d^\m_j (H^{j a i} - \Ht^{j a i}) \F_{a i},\qquad \m = 0,1,\cdots,5.
\ee
It is clear from the structure of \eqref{Noecurr} that the Noether charge $Q=\int d^5x\,j^0$ is identically zero off-shell
if the temporal direction is in the 2d subspace of the `2+4' dimensional space-time.
Therefore, in this formulation we lose the freedom to place the time direction in the 4d subspace. This makes the 2+4 splitting different from the 1+5 and 3+3 splittings of the previous formulations of the 6d chiral 2-form action.

The field equations  which one obtains by varying \eqref{quadraticS} are
\bea
\del_k \lrbrk{ -\Ht^{a ki} + H^{a k i} } &=& 0,  \label{Leom1}\\
\del_k \lrbrk{ -\Ht^{ijk}
+ 2H^{ijk} } + \del_a H^{a ij} &=& 0.  \label{Leom2}
\eea
Equation \eqref{Leom1} has the general solution
\be \label{Lgsoln1}
-\Ht^{a ik} + H^{a ik} = \e^{ab}\e^{ikjl} \del_j \tilde\F_{b l},
\ee
where $\tilde\F_{b l}$ satisfy the condition \eqref{paraCond},
because the left-hand-side of the above equation is anti-self-dual.
Hence, we can obtain the self-duality equation
\be \label{LSD1}
H_{a ij} = \Ht_{a ij}
\ee
by fixing the semi-local gauge symmetry \eqref{LSMsym} appropriately.
Substituting \eqref{LSD1} into \eqref{Leom2} and using the Bianchi identity,
we get
\be
\pa_k \lrbrk{ -\Ht^{ijk} + H^{ijk} } = 0,
\ee
which has the general solution
\be
-\Ht_{ijk} + H_{ijk} = \ove{2}\e_{ab}\e_{ijkl}\del^l \tilde\Omega^{ab},
\ee
where $\tilde\Omega_{ab}$ are arbitrary functions which can be put to zero with the use of the local gauge transformations \eqref{Lsym1}.
We thus arrive at another set of self-duality equations
\be\label{LSD2}
H_{ijk} = \Ht_{ijk}.
\ee
Combined together, eqs. \eqref{LSD1} and \eqref{LSD2} are equivalent to \eqref{H=*H}.

The action \eqref{quadraticS} is manifestly invariant under
$SO(1,1)\times SO(4)$
subgroup of Lorentz symmetry.
However, it is less obvious that
the action also enjoys the modified Lorentz symmetry
with parameters $\l^a{}_j \equiv \l^a_{\color{black}j}$ ($\l_a{}^i \equiv {\color{black}\l_a^i}$) associated with the coset transformations
$SO(1,5)/[SO(1,1)\times SO(4)]$.
For simplicity, we present the modified part of the $SO(1,5)$ Lorentz symmetry
in the gauge $B_{ab} = 0$
\be \label{LMLsym}
\d B_{a i} = \d_1 B_{a i} + \d_2 B_{a i}, \qquad \d B_{ij} = \d_1 B_{ij} + \d_2 B_{ij},
\ee
with
\bea \label{stdLorentz}
\d_1 B_{a i} &=& \l_a^j B_{ji} + \l^b_j(x_b\del^j - x^j\del_b)B_{a i},   \nn\\
\d_1 B_{ij} &=& - \l^b_i B_{b j} + \l^b_j B_{b i} +  \l^b_k(x_b\del^k - x^k\del_b)B_{ij},
\eea
being the standard Lorentz transformation
and\footnote{\label{ft:d3B}There is a room of adding to this transformation another term
\be \label{LmodLorentz3}
 \d_3 B_{a i} = \l_b^j x^b (H-\Ht)_{a ij}.
\ee
One may check that under this transformation the Lagrangian is invariant
up to a total derivative term
\be \label{d3L}
\delta_3 S = \ove{2}\int d^6x \partial_k
(\l_b^j x^b {\mathcal H}_{a ij} {\mathcal H}^{aik}).
\ee
}
\be \label{LmodLorentz}
 \d_2 B_{a i} = \l^b_j x^j (H-\Ht)_{ba i}, \qquad \d_2 B_{ij} = \ove{2}\l^b_k x^k (H-\Ht)_{b ij}
\ee
vanish on the mass shell. Thus, the modified $SO(1,5)$ Lorentz symmetry reduces to the standard one when the field strength of the 2-form $B_2$ is self-dual.

\subsection{Action for the Abelian $\mathcal N=(2,0)$ tensor supermultiplet} \label{sec:susy}
The supersymmetric extension of the free chiral 2--form action is obtained by adding to it kinetic terms for five scalar fields $X^I(x)$ ($I=1,2,3,4,5$) and a sixteen--component fermionic field $\psi(x)$ which together with $B_{\mu\nu}(x)$ form an $\mathcal N=(2,0)$ supermultiplet. The resulting action
\be\label{n20}
S = \ove{2}\int_{\cM_6} d^6x \lrsbrk{
-\lrbrk{ \hlf\Ht_{ab i}H^{ab i} + \ove{4}H_{a ij}H^{a ij} + \ove{6}H_{ijk}H^{ijk}}
	+ \lrbrk{ i\bar\psi \G^\m \pa_\m \psi - \pa_\m X^I \pa^\m X^I }
	}
\ee
is invariant under the following $\mathcal N=(2,0)$ supersymmetry transformations with a sixteen--component constant spinor parameter $\e$
\bea\label{susy}
\d X^I &=& i\bar\e \G^I \psi, \nn\\
\d B_{\m\n} &=& i\bar\e \G_{\m\n} \psi, \nn\\
\d \psi &=& \G^\m \G^I \pa_{\m} X^I \e + \ove{12}\G_{\m\n\r}K^{\m\n\r} \e,
\eea
where the self-dual 3-form $K={}^*K$ is defined in terms of the components of the field strength $H_{\mu\nu\rho}$ as follows
\be
K^{\m\n\r} = \frac 12(H^{\m\n\r}+\tilde H^{\m\n\r}) +\frac 12(H^{ijk}-\tilde H^{ijk})  \d_i^{\m}\d_j^\n \d^\r_k - \frac 32 (H^{abj}-\tilde H^{abj}) \d_a^{[\m} \d_b^\n \d^{\r]}_j.
\ee
Note that the self--dual Lagrangian \eqref{quadraticS} for $H_3$ is equal to $L=\frac 16 H_{\mu\nu\rho} K^{\m\n\r} $.

To define the sixteen--component spinors we use the  same conventions as in the Appendix of \cite{Lambert:2010wm}. Namely,
$\Gamma^\mu$ $(\mu=0,1,2,3,4,5)$ and $\Gamma^I$ $(I=6,7,8,9,10)$ are the $32\times 32$ $D=11$ gamma--matrices in the Majorana representation, and the 32--component Majorana spinors $\psi(x)$ and $\e$ are subject to the chirality constraints
\be\label{chirality}
\psi=-\g^\6\psi,\qquad \e=\g^\6\e, \qquad \g^\6 = \ove{6!}\e^{\m_1\cdots\m_6} \G_{\m_1\cdots\m_6}
\ee
which reduce the number of the independent spinor components down to sixteen.

To study  the theory described by the action \eqref{n20}, in particular its topological twisting \cite{Anderson:2013hpa,Gran:2014lia}, in geometrically non--trivial 6d backgrounds, one should couple the action \eqref{n20} to a $d=6$ supergravity in a way similar to that considered in the `1+5' formulation \cite{Dall'Agata:1997db,Riccioni:1998pj,Riccioni:1999xq,Riccioni:2001bg,VanHoof:1999xi} and to look for $d=6$ backgrounds preserving at least part of supersymmetry.
This is left as a future work.

\setcounter{equation}0
\section{Non-linear chiral 2-form gauge theory with non-manifest 6d Lorentz-invariance} \label{sec:nonlinear}
We would like to find a non-linear generalization of the action \eqref{quadraticS}
in 6d Minkowski space, with the ultimate aim to describe the M5-brane.
Let us stress that we are deforming the free theory \eqref{quadraticS}
to a nonlinear one in a six dimensional Minkowski space.
An attempt to apply the construction to the M5-brane worldvolume theory with a non--trival induced 6d metric will be discussed in section \ref{sec:iss}.
To have a hint on how this generalization should be carried out, let us first rewrite the quadratic action in the following form
\be\label{quadraticS1}
S =
	- \int d^6x\lrbrk{\hlf\Ht_{ab i}H^{ab i} + \ove{2}H^-_{a ij}H^{+a ij} + \ove{6}H_{ijk}H^{ijk}},
\ee
in which the second term is the product of the anti-self-dual \eqref{-} and self-dual \eqref{+} part of $H_{aij}$.

The non-linear action we are interested in should respect the same (possibly non-linearly modified) symmetries of the free chiral field action and should produce the equations of motion, in particular, the same non-linear self-duality condition on $H_3$ which follow
from the other formulations of the M5-brane dynamics, namely, from the superembedding description \cite{Howe:1996yn,Howe:1997fb} and the M5-brane actions \cite{Bandos:1997ui,Aganagic:1997zq,Ko:2013dka}
when they are put in a Minkowski target superspace with the M5--brane excitations along the transverse directions frozen.

We have found that the non-linear 6d action which satisfies these requirements has the following form
\be\label{S24}
S = 
	- \int d^6x
	\lrbrk{\hlf\Ht_{ab i}H^{ab i} + \ove{2}H^-_{a ij}H^{+a ij} + \cI(H_{ijk}, H^+_{aij}) },
\ee
where $\cI$ is the following functional of $H_{ijk}$ and $H^+_{aij}$ only
\be \label{cI}
\cI(H_{ijk}, H^+_{aij}) = G^2 + 4\frac{Q-1}{Q}\sqrt{1+G^2-\frac{Q^2X}{64}}.
\ee
In \eqref{cI}
\be \label{Gdef}
G_{\color{black}l} = \ove{3!}\e_{ijkl}H^{ijk}, \qquad
\ee
and later one we will also deal with
\be
\Gt_{\color{black}l} = \ove{3!}\e_{ijkl} \Ht^{ijk},
\ee
{\color{black}where the Hodge dual field strength $\Ht$
is taken with respect to Minkowski metric as in \eqref{flatDual}.}
$X,Y$ and $G^2$ are three $SO(1,1)\times SO(4)$ invariant scalars
\be\label{X}
X = -2H^+_{aij}H^+_{b}{}^{ij}H^{+akl}H^{+b}{}_{kl}, \quad
Y = -2G^kG_i H^+_{a jk} H^{+a ij}, \quad
G^2 = G^iG_i,
\ee
and
$Q$ satisfies the following sextic equation
\bea \label{sextic}
&&-16(G^2)^3 - 16(G^2)^2
+16Q \left((G^2)^3+(G^2)^2+G^2 Y+Y\right)  \nn\\
&& +Q^2 \left( (G^2)^4+\frac{(G^2)^2 X}{4} - 2(G^2)^2 Y
		     +G^2 X - 16G^2 Y + X + Y^2 - 16Y \right)  \nn\\
&& +Q^3 \left(-\frac{(G^2)^2 X}{2} - G^2 X - X \right)
+\ove{8} Q^4 \left( (G^2)^2 X + X Y \right)
+\frac{Q^6 X^2}{256}
= 0.
\eea
Using exactly the same analysis as in Section \ref{sec:free},
one can show that the variation of the action \eqref{S24} leads to the following non-linear self-duality equations
\be\label{nos}
H^-_{a ij} = -\ove{4} \frac{\pa \cI}{\pa H^{+a ij}}, \qquad
\Gt_i = \ove{2} \frac{\pa \cI}{\pa G^i} \quad \leftrightarrow \quad
H_{ab{\color{black}i}}=\ove{\color{black}12} \varepsilon_{abijkl}\frac{\pa \cI}{\pa H_{jkl}},
\ee
which are equivalent to those obtained from the superembedding description of the M5--brane \cite{Howe:1996yn,Howe:1997fb}, as we will show below and in Appendix A.

\subsection{Construction of the non-linear action in 6d Minkowski space}
Let us first consider \eqref{S24} with, a priori, unknown generic functional $\cI(H_{ijk},H^+_{alm})$,
and require the action to be invariant under the modified Lorentz symmetry \eqref{LMLsym},
which is composed of the standard Lorentz transformations \eqref{stdLorentz}
and the additional terms
\be
\d_2 B_{a i} = \e_{ab}\l^b_j x^j \lrbrk{ -\Gt_i + \ove{2}\frac{\pa \cI}{\pa G^i} },
\ee
\be
\d_2 B_{ij} =
	\l^b_k x^k \lrbrk{  H^-_{b ij} + \ove{4} \frac{\pa \cI}{\pa H^{+b ij}} },
\ee
with the parameter $\lambda^b_{\color{black}k}$ taking values in $SO(1,5)/[SO(1,1)\times SO(4)]$.\footnote{Just as in the linear theory, there is an ambiguity in the transformation rule of $B_2$. The following variation
\be
\d_3 B_{a i} = \l_b^j x^b \lrbrk{H^-_{\a ij} + \ove{4}\frac{\pa\cI}{\pa H^{+aij}}}
\ee
also leaves the Lagrangian invariant up to total derivative terms. { The nature and meaning of this ambiguity is unclear to us.}}
The above transformations reduce to \eqref{LmodLorentz} if
we put $\cI = G^2$.
Notice also that on the mass shell \eqref{nos} the modified Lorentz symmetry reduces to the standard one.

After a somewhat lengthy calculation, the requirement of the invariance of the action \eqref{S24}  under the modified Lorentz transformations leads to the following constraint on the form of $\cI$
\be \label{ncSC}
-2G_i H^{+a ij}
+ \frac{\pa\cI}{\pa G^i} H^{+a ij}
+ \ove{2}  G_i \lrbrk{\frac{\pa \cI}{\pa H^+}}^{a ij}
+ \ove{4} \lrbrk{\frac{\pa \cI}{\pa G}}_i \lrbrk{\frac{\pa \cI}{\pa H^+}}^{a ij} = 0.
\ee
Note that the symmetry constraint is trivially satisfied if we have $\cI = G^2$.

As is well known (see e.g. \cite{Gaillard:1981rj,Gibbons:1995cv,Perry:1996mk,Bossard:2011ij,Carrasco:2011jv,Pasti:2012wv}), the above constraint may have different solutions corresponding to different nonlinear chiral 2-form
theories.
To fix the form of $\cI$, our strategy will be to first find the action which leads to the self-duality equations which are equivalent to the ones given by the superembedding formulation of the M5 brane \cite{Howe:1996yn,Howe:1997fb},
and then check that the solution satisfies the constraint \eqref{ncSC}.

In the superembedding description of the M5-brane \cite{Howe:1996yn,Howe:1997fb} the field strength $H_3$ of the chiral field $B_2$ is expressed in terms of a (linear) self-dual tensor $h_3=*h_3$ as follows\footnote{Our normalization of the field strengths differs from that in \cite{Howe:1997vn} by the factor of $\frac 14$ in front of $H_3$.}
\be\label{embsd}
\frac 14 H_{\m\n\rho}=m^{-1\lambda}_{\mu}h_{\lambda\n\rho}\,, \qquad \frac 14\tilde H^{\m_1\n_1\rho_1}=\frac{1}6\epsilon^{\m_1\n_1\rho_1\m\n\rho}m^{-1\lambda}_{\mu}h_{\lambda\n\rho}=Q^{-1}m^{\mu_1\lambda}
h_{\lambda}{}^{\n_1\rho_1}\,
\ee
where $m^{-1\lambda}_{\mu}$ is the inverse matrix of
\be\label{mk}
m_{\mu}{}^{\lambda}=\delta_{\mu}{}^{\lambda}-2k_{\mu}{}^{\lambda}\,,\qquad m_\mu^{-1\lambda}=Q^{-1}(2\delta_{\mu}{}^{\lambda}-m_{\mu}{}^{\lambda}),\qquad k_{\mu}{}^{\lambda}=h_{\mu\nu\rho}h^{\lambda\nu\rho}\,
\ee
and
\be\label{Q}
Q=1-\frac 23 \tr \,k^2\,.
\ee

As shown in \cite{Howe:1997vn, Ko:2013dka}, by splitting the six-dimensional indices into 1+5 or 3+3,
one can derive the duality equations which are produced by the M5-brane actions in the corresponding formalisms. Here we perform the
`2+4' splitting of \eqref{embsd}
and obtain
\be \label{SEH+}
\ove{4} H^+_{a ij} = Q^{-1}h_{a ij},
\ee
\be \label{SEH-}
\ove{4} H^-_{a ij} = Q^{-1}\lrsbrk{4g^2 h_{a ij} + 8g^mg_ih_{a jm} - 8g^m g_j h_{a im} + 2h_{a xy}h^{b xy}h_{b ij}},
\ee
\be \label{SEG}
\ove{4} G^l = Q^{-1}\lrsbrk{(1+4g^2)g^l - 4 g^x h_{a xk}h^{a lk} },
\ee
\be \label{SEGt}
\ove{4}\Gt^l = Q^{-1}\lrsbrk{(1-4g^2)g^l + 4 g^x h_{a xk}h^{a lk} },
\ee
\be  \label{SEQ}
Q = 1 - 16(g^2)^2 - 2h_{a ij}h^{b ij}h_{b kl}h^{a kl} + 32g^l g_k h^{a jk}h_{a jl},
\ee
where\footnote{ The reader should not confuse $g^2$ with the square of the determinant of the $6d$ metric $g=\det g_{\m\n}$. }
\be
g_k \equiv \ove{3!}\e_{lijk}h^{lij}, \qquad g^2 \equiv g_k g^k,
\ee
with $a,b=0,5$ and $i,j,\cdots = 1,2,3,4$.
By expressing $g_i$ and $h_{a ij}$ in \eqref{SEH-} and \eqref{SEGt} in terms of
$H^+_{a ij}$ and $G_i$ via \eqref{SEH+} and \eqref{SEG}, one may, in principle, obtain the desired duality equations.
However, it is very difficult to proceed directly in this way.

So, let us present a heuristic way to obtain the action.
In \cite{Ko:2013dka}  it was found that the chiral 2-form parts of both, the 3+3 and  1+5  formulation of the M5-brane action have the same on-shell value determined by the super-embedding scalar $Q$ \eqref{Q}, namely
\be \label{osS}
S^{\text{on-shell}} = - \int d^6x\sqrt{-g} \, \frac{4}{Q}.
\ee
We assume that this is also true in the 2+4 formulation.
Thus, with the help of the superembedding equations \eqref{SEH+}-\eqref{SEQ} we first rewrite the Lagrangian as follows
\be
\hlf\Ht_{ab i}H^{ab i} + \ove{4}H_{a ij}H^{a ij} + \cI
= -G^i\Gt_i + 8\frac{1-Q}{Q^2} - \ove{2}(G^2-\Gt^2) + \cI.
\ee

The on-shell action \eqref{S24} equals \eqref{osS} (in the
6d Minkowski space) if
\be \label{OScI}
\cI_\text{on-shell} = 4(3Q^{-1}-2Q^{-2}) + G^i\Gt_i + \ove{2}(G^iG_i - \Gt^i\Gt_i).
\ee
To recover the off-shell action $\cI$, one needs to replace
the terms with $\Gt_i$ by $H_{a ij}$ and $G_i$.

It is convenient to rewrite \eqref{SEQ} as
\be \label{Qinxyg}
Q = 1 - 16(g^2)^2 + 16y + x,
\ee
where
\be\label{xy}
x \equiv -2h_{a ij}h_b{}^{ij}h^{a kl}h^b{}_{kl}, \quad
y \equiv -2g^k g_i h_{a jk}h^{a ij}.
\ee
Note that
\be \label{xX}
256x=Q^4X,
\ee
where $X$ was defined in \eqref{X}.
From the superembedding equations and \eqref{Qinxyg} we obtain
\be \label{G2se}
\ove{16}G^2 =Q^{-2}\lrsbrk{g^2(2+4g^2-Q) + \frac{1-Q+x}{4} },
\ee
\be  \label{Gt2se}
\ove{16}\Gt^2 =Q^{-2}\lrsbrk{g^2(2-4g^2-Q) - \frac{1-Q+x}{4} },
\ee
\be
\ove{16}G^i\Gt_i = g^2 Q^{-1}.
\ee
We can solve $g^2$ in terms of $X,Y,G^2,Q$ from \eqref{G2se} together with \eqref{xX},
\be \label{g2inHG}
g^2 = \ove{8} \lrbrk{ -2+Q + \sqrt{Q^2 + G^2Q^2 - Q^4X/64} }.
\ee
Upon inserting all these ingredients into \eqref{OScI}, we find \eqref{cI}.
The remaining work is then to derive the equation satisfied by $Q$.
We start by expressing $Y$, defined in \eqref{X}, in terms of $h_3$ using the superembedding equations
\be \label{Yinxyg}
\ove{256}Y
= Q^{-4}\lrsbrk{ y - xy +4(g^2)^2(x+4y)+g^2(x+8y) }.
\ee
Using \eqref{Qinxyg}, \eqref{xX} and \eqref{g2inHG} in the above equation,
one obtains the following expression for the square root
$
 \sqrt{1 + G^2 - Q^2X/64},
$
\be
 \sqrt{1 + G^2 - Q^2X/64}
 = \frac{- (G^2)^2 Q - 8 G^2 (Q-1)+Q^3 X/16+ Q (Y-8)+8}{4 ({G^2}+2) (Q-1)}.
\ee
The polynomial equation obtained from the above relation  is
\eqref{sextic}.
As a consistency check, one can substitute \eqref{Qinxyg}, \eqref{xX},
\eqref{G2se} and \eqref{Yinxyg} into the sextic equation
to see that it is trivially satisfied.

As one can show (see Appendix), the resulting $\cI$ constructed in terms of $X,Y,G^2$ and $Q$ as in \eqref{cI} satisfies \eqref{ncSC} and hence
is the promising candidate for the non-linear off-shell action we are looking for.

However, one should still show that the self-duality equations obtained from this action
are equivalent to those in the superembedding formulation. The detailed proof of this is given in the Appendix A.

Solutions to a generic sextic equation can be written in terms of the Kamp\'{e} de F\'{e}riet hypergeometric function. Only some special sextic equations can be factorized into
radicals and hence solved explicitly \cite{sextic},
however, this is not the case for \eqref{sextic}.

Only when $Y=G^2=0$, \eqref{sextic} reduces to
\be
Q^2X \lrbrk{ \frac{Q^4X}{256} - Q + 1 } = 0.
\ee
Although our sextic equation is not exactly solvable, we can always reconstruct its solution perturbatively as a series in powers of the field strengths.

\setcounter{equation}0
\section{Attempts to get a full M5-brane action and issues \label{sec:iss}}
As our goal is to provide another alternative action for the M-theory five-brane,
we will look for the covariantisation of the actions \eqref{quadraticS} and \eqref{S24}.
However, as we will find out in this section, putting the actions \eqref{quadraticS} and \eqref{S24}
on a curved
6d background turns out to be highly nontrivial.
The standard PST technique is not straightforwardly applicable in this case.
We will first present an incomplete covariantisation of the action \eqref{quadraticS},
in which the theory is \textit{formally} covariant yet not fully consistent.
Nevertheless, Hamiltonian analysis suggests that coupling to gravity looks promising.

\subsection{Covariantisation issue \label{sec:cov}}
In this section, we present an attempt to covariantise
the 2+4 chiral 2-form action and couple it to 6d gravity
\eqref{quadraticS}
using the standard PST technique.
Unlike \cite{Pasti:1996vs,Pasti:2009xc},
the result of the covariantisation for 2+4
turns out to be problematic in that
the  action does not acquire the PST gauge symmetry and, hence, auxiliary PST scalars carry undesirable dynamical degrees of freedom.
We will discuss these issues in detail in section \ref{sec:dis}.

Let us introduce a doublet of auxiliary scalar fields \footnote{One can alternatively choose a quadruplet of auxiliary fields,
$a^s(x)$, $s=1,2,3,4$. In that case, they label a 4D representation of an internal
$GL(4)$ symmetry.}
$a^s(x)$ with $s= 1,2$ labelling a
2D representation of internal
rigid $GL(2)$ symmetry of the action. Using the derivatives of $a^s$ we construct the projector matrices
\be
P_\m{}^\n = \pa_\mu a^r Y^{-1}_{rs} \pa^\n a^s, \quad \Pi_\m{}^\n = \d_\m^\n - P_\m{}^\n, \quad \P_\m{}^\n \pa_\n a^s = 0,
\ee
where
\be
Y^{rs} \equiv \pa_\rho a^r \pa_\sigma a^s g^{\rho\sigma}(x),
\ee
and $g^{\rho\sigma}(x)$ is the inverse of the 6d metric $g_{\mu\nu}(x)$.
The projector  $P_\m{}^\n$ has rank 2 and $\P_\m{}^\n$ has rank 4, so they
split the 6d directions into 2+4 ones which are orthogonal to each other.

The projectors satisfy the following identities
\be
3\e^{\m\n\r\t\s\l} P_\m^{[\k} P_\n^{\eta} \P_\r^{\xi]} = - \e^{\m\n\r\k\eta\xi} \P_\m^{[\t} \P_\n^{\s} \P_\r^{\l]}, \quad
\e^{\m\n\r\t\s\l} P_\m^{[\k} \P_\n^{\eta} \P_\r^{\xi]} = - \e^{\m\n\r\k\eta\xi} P_\m^{[\t} \P_\n^{\s} \P_\r^{\l]},
\ee
\be \label{dP}
\Pi_{[\rho}{}^\lambda\Pi_{\kappa]}{}^\mu D_{\lambda}P_{\mu}^\nu=0=\Pi_{[\rho}{}^\lambda\Pi_{\kappa]}{}^\mu D_{\lambda}\Pi_{\mu}^\nu\,\,
\ee
where $D_\mu$ is the covariant derivative associated with the metric $g_{\mu\nu}$.

The proposed ansatz for the covariantised form of the action \eqref{quadraticS} is
\be \label{LcovS}
S = - \int d^6x \sqrt{-g} \lrsbrk{ \ove{2} (\Ht PP\P H)
	+ \ove{4} (HP\P\P H)
	+ \ove{6} (H\P\P\P H)  },
\ee
where $g$ is the determinant of the 6d metric,
the dual field strength is
\be
\Ht^{\r\m\n} = \ove{6\sqrt{-g}} \e^{\r\m\n\l\s\t} H_{\l\s\t},
\ee
and
\be
(\Ht PP\P H) = \Ht_{\m\n\r} H^{\t\s\l} P_\t^\m P_\s^\n \P_\l^\r, \quad
(HP\P\P H) = H_{\m\n\r} H^{\t\s\l} P^{\m}_\t \P^\n_\s \P^\r_\l,   \nn
\ee
\be
(H\P\P\P H) = H_{\m\n\r} H^{\t\s\l} \P_\t^\m \P_\s^\n \P_\l^\r.
\ee
The action enjoys the covariant versions of the local gauge symmetry
\be \label{LCasym}
\d B_{\m\n} = P^\r_{[\m} P^\s_{\n]} \Omega_{\r\s}, \qquad \d a^s =0,
\ee
as well as the semi-local gauge symmetry
\be \label{LSMsymC}
\d B_{\m\n} = P_{[\m}^\r \P^\s_{\n]} \F_{\r\s}, \qquad \d a^s = 0,
\ee
with $\F_{\r\s}$ satisfying the anti-self-duality condition
\be \label{LCsmcond}
\delta H_{\r\s\l}\, P^\r_{\m} \P^\s_{\n} \P^\l_{\r} = -\delta {\tilde H}_{\r\s\l}\, P^\r_{\m} \P^\s_{\n} \P^\l_{\r}, \qquad \delta H_{\r\s\l}= 3\partial_{[\lambda} (P_{\r}^\kappa \P^\delta_{\s]} \F_{\kappa\delta})\,\,.
\ee
The Noether current associated with this symmetry is
\be\label{j}
j^\m = (H^-P\P\P)^{\r\s\m} (\F P\P)_{\r\s},
\ee
where we introduced the anti-self-dual part of the field strength
\be\label{-}
H^-_{\m\n\r} \equiv \ove{2} \lrbrk{H_{\m\n\r} - \Ht_{\m\n\r}}, \qquad
\ee
the corresponding self-dual part being
\be\label{+}
H^+_{\m\n\r} \equiv \ove{2} \lrbrk{H_{\m\n\r} + \Ht_{\m\n\r}}.
\ee
It is clear that the Noether charge associated with \eqref{j} vanishes {\textit{off-shell}} if we align the time
along the directions singled out by the $P$-projector, i.e. along the `2'-subspace of `2+4'. Therefore, \eqref{LSMsymC} is eligible to be a gauge symmetry,
and we can use it to obtain the self-duality equations.

The field equation obtained as the result of the variation of
the {\color{black}formally} covariant action with respect to $B_{\mu\nu}$ is
\be \label{LCeom}
\del_{\r} \lrsbrk{ \sqrt{-g}
\Big( 6(H^- P\P\P)^{[\m\n\r]} + 4(H^-\P\P\P )^{[\m\n\r]} \Big) } = 0.
\ee
Its integration gives
\be \label{gensoln24}
\color{black}
 \sqrt{-g}
 \Big( 6(H^- P\P\P)^{[\m\n\r]} + 4(H^-\P\P\P )^{[\m\n\r]}  \Big)
 = \e^{\m\n\r\t\s\l}
 \del_{\t} \lrbrk{ \tilde\F_{\k\eta} P^{\k}_{\s} \P^{\eta}_{\l} +  \tilde\Omega_{\k\eta} P^{\k}_{\s} P^{\eta}_{\l} },
\ee
for some parameters $\tilde\F$ and $\tilde\Omega$.
Projecting
both sides of the above equation on $P\P\P$,
we get
\be\label{LCprojgsoln}
6\sqrt{-g} (H^-P\P\P)^{[\m\n\r]}
= 3 \e^{\t\s\l \k\eta\xi} \del_\k (\tilde\F P\P)_{\eta\xi}
	P^{[\m}_\t \P^\n_\s \P^{\r]}_\l.
\ee
Notice that $\tilde\F_{\m\n}$ satisfy the constraint \eqref{LCsmcond},
because the left-hand-side  of \eqref{LCprojgsoln} is anti-self-dual.
Thus, by appropriately fixing  the gauge symmetry \eqref{LSMsymC},
we get the first set of the duality equations
\be\label{-ppp}
(H^- P\P\P)^{[\m\n\r]} = 0.
\ee
Substituting this back into the equation \eqref{LCeom},
we obtain
\be
4\sqrt{-g} (H^-\P\P\P)^{[\m\n\r]} = \e^{\m\n\r \t\s\l} \del_{\t} \lrbrk{ \tilde\Omega PP}_{\s\l}.
\ee
The appropriate choice of the gauge symmetry \eqref{LCasym}
leads to the other set of duality equations
\be\label{-pppp}
(H^-\P\P\P)^{[\m\n\r]} = 0.
\ee
The equations \eqref{-ppp} and \eqref{-pppp}
amount to the self-duality of the field strength $H_3$.

The crucial ingredient of the PST covariantisation technique is that,
in addition to the gauge symmetries \eqref{LCasym} and \eqref{LSMsymC},
the action
should be also invariant under the
PST gauge symmetry, in which the auxiliary fields
$a^s(x)$ transform by arbitrary local functions.
In view of \eqref{LmodLorentz},
a reasonable guess for the PST gauge transformation
would be
\footnote{It turns out that this transformation does not leave the action invariant.
One might try to add to the transformation law a term $\d B_{\m\n}
\ni V^\rho(x)(H^-P\P\P)_{\r\m\n}$, where
$V^\rho$ is gauge-fixed to be $\l_b^jx^b$
if $a^s=\d^s_ax^a$ is allowed,
motivated by the Footnote \ref{ft:d3B}.
However, this turns out to be not helpful.}
\bea \label{LPST24}
\d a^s = \vphi^s, \quad
\d B_{\m\n} = 6\L^\r (H^- PP\P)_{[\m\n\r]}
			+ 3 \L^{\r} (H^- P\P\P)_{[\m\n\r]},
\eea
where
\be
\L^\r \equiv \vphi^s Y^{-1}_{st}\del^\r a^t.
\ee
{\color{black}
If the above transformation were indeed a gauge symmetry},
one could gauge fix the auxiliary fields $a^s(x)$ to
coincide with two (worldvolume)
coordinates
$x^a$ ($a=0,5$), thus obtaining (in the flat worldvolume space)
the non-manifestly Lorentz invariant action of the previous section
\be
a^s = \d^s_a x^a.
\ee
This gauge-fixing condition would be preserved by a combined
Lorentz transformation with parameter $\l^a_i$
and the transformation \eqref{LPST24},
with parameter $\L^\r=-\d^\r_b \l^b_j x^j$,
\be
\lrbrk{\d_\text{Lorentz} + \d_\text{PST}}a^s = 0.
\ee
This combination of two transformations acting on the chiral 2-forms would give exactly the modified Lorentz symmetry \eqref{LMLsym}.

However, we find that \eqref{LPST24} leaves the action
invariant, up to total derivative terms,
only when the following constraints are satisfied,
\be \label{PSTcond24}
P^{\r}_{\m}P^\s_{\n}D_{(\r} \L_{\s)} = 0 = \P^{\r}_{\m} \P^{\s}_{\n} D_{(\r} \L_{\s)}.
\ee
Therefore, the proposed transformation \eqref{LPST24}
is not eligible to be a PST \emph{gauge} symmetry of the action \eqref{LcovS}.
The failure of PST gauge symmetry implies the inconsistency
of the current covariantisation procedure.
Together with other issues we will discuss in more detail in section \ref{sec:dis},
this indicates the trouble with coupling of the 2+4 formulation to 6d gravity.

\subsection{Discussion of the issues}\label{sec:dis}
We have seen in section \ref{sec:cov} that the standard PST covariantisation
is not  applicable (at least straightforwardly) to the chiral 2-form theory with the 2+4 splitting of six dimensions.
In this section, we will study the encountered issues in more detail with the hope of
understanding the origin of the problems and resolving them in future.

\subsubsection*{Failure of finding PST gauge symmetry}
The PST gauge transformation \eqref{LPST24} leaves the action \eqref{LcovS} invariant
only when the constraints \eqref{PSTcond24} are satisfied.
Usually, a fully-fledged PST gauge transformation allows us to gauge-fix
the auxiliary fields, say $a^s=\d^s_ax^a$, so that the covariant theory reduces
to the non-manifestly covariant one.
Obviously, the constraints \eqref{PSTcond24} set obstacles to do this.
In the absence of the PST gauge symmetry in the formulation of section \ref{sec:cov}, the fields $a^s$ are not really auxiliary and may carry undesirable dynamical degrees of freedom, as the following analysis shows.

\subsubsection*{(In)dependence of the field equations of $a^s(x)$}
In the free theory \eqref{LcovS}, the field equation of the 2-form gauge field
derived from the action principle is
\be \label{LCeom}
\del_{\r} \lrsbrk{ \sqrt{-g}
\Big( 6(H^- P\P\P)^{[\m\n\r]} + 4(H^-\P\P\P )^{[\m\n\r]} \Big) } = 0.
\ee
On the other hand, the field equations of the fields $a^s$ are
\be \label{Leoma}
\pa^\s \lrsbrk{{\color{black}\sqrt{-g}} Y^{-1}_{st}\pa_\r a^t (H^-\P\P\P)_{[\m\n\s]} (H^- P\P\P)^{[\r\m\n]} } = 0.
\ee
If $a^s(x)$ were really auxiliary, their field equations would not be independent
but implied by the \textit{second order} field equation of the 2-form gauge field.
This is the case when the PST covariantisation is successful as
in \cite{Pasti:1996vs,Pasti:2009xc}.
Actually, the existence of the PST gauge symmetry in \cite{Pasti:1996vs,Pasti:2009xc}
is guaranteed by
the fact that the field equations of the auxiliary field(s) are redundant.
In our 2+4 splitting case, however, one can readily verify that \eqref{Leoma} is not implied by the second order field equation \eqref{LCeom} and, hence the fields $a^s$ may actually carry additional
dynamical degrees of freedom.

\subsubsection*{Issue with modified diffeomorphism}
The chiral 2-form actions of \cite{Pasti:1996vs,Pasti:2009xc} are manifestly 6d diffeomorphism invariant.
Upon the appropriate gauge fixing of the auxiliary fields,
one can obtain the non-manifestly reparametrization invariant actions.
Such  actions
are invariant under certain modified diffeomorphism transformations,
which reduce to the standard ones on-shell.
For example, \cite{Schwarz:1997mc} is such a theory which is nonlinear in the gauge field.

For simplicity, we will consider only free theories here,
as the issue of covariantisation in 2+4 splitting arises already therein.
Let us now review how the non-manifest diffeomorphism invariance works in the theories with
 1+5 and 3+3 splitting and point out the issue with the 2+4 splitting model.

The following action based on the 1+5 splitting of six dimensions
($m,n,p,q,k=0,1,2,3,4$)
\be \label{15diffS}
S = \int d^6x \lrbrk{ \ove{4}\Ht^{5mn}H_{5mn}   + \cI },
\ee
\be
\cI = + \ove{8}\e_{mnk5pq} \Ht^{5mn} \Ht^{5pq}\frac{g^{5k}}{g^{55}}
	- \ove{4\sqrt{-g}} \Ht^{5pq} \Ht^{5mn} \ove{g^{55}} g_{pm} g_{qn},
\ee
is the truncation of the nonlinear theory \cite{Schwarz:1997mc} to the linear order.
The $\Ht^{5mn}$ is defined without involving any metric
\be
\Ht^{5mn} \equiv \ove{3!}\e^{5mnpqk}H_{pqk}.
\ee
One can alternatively formulate a 1+5 theory by singling out the temporal direction from other five
spatial ones. In this case, the resulting action is  the Henneaux-Teitelboim 1+5 (HT) action \cite{Henneaux:1987hz,Henneaux:1988gg}.

Though it is not obvious, this action also has the modified diffeomorphism symmetry
\be \label{15diff}
\d B_{mn} = -\xi \frac{\pa \cI}{\pa \Ht^{mn5}},
\ee
with the diffeomorphism parameter in the fifth spacial direction $x^5 \rar x^5+ \xi$,
as well as the standard diffeomorphism for $x^k \rar x^k + \xi^k$ $(k=0,1,2,3,4)$.
The transformation law \eqref{15diff} reduces to the standard one on-shell.
The components $B_{k5}$ do not transform because we work in the gauge $B_{k5}=0$ for simplicity, since $B_{k5}$ enters the action through a total derivative term.

A non-manifestly diffeomorphism invariant action can also be obtained in the formulation with the 3+3 splitting \cite{Pasti:2009xc,Ko:2013dka}
by gauge fixing values of the triplet of auxiliary fields $a^s = \d^s_a x^a$, $s=1,2,3$
( in the following $a,b,c,d=0,1,2$, $i,j,k,l,m,n, p,q=3,4,5$)
\be \label{33diffS}
S = \int d^6x \lrbrk{ \ove{36} \e^{abc}H_{abc} \e^{ijk}H_{ijk}
	- \ove{{\color{black}4}} \e^{abc}\e^{ijk} H_{bc k} H_{a ij}
	+ \cI_1 + \cI_2},
\ee
\be
\cI_1 = - \e^{ijk}\e^{abc} F^l_c G \frac{g_{i a}g_{j b} g_{kl} }{\det(g_{mn})}
	+ \ove{3}\e^{ijk}\e^{abc} G^2 \frac{g_{a i}g_{b j}g_{c k}}{\det(g_{mn})}
	- \e_{ijk}\e^{abc} F^i_a F^j_b g^{-1}_{cd} g^{d k},
\ee
\be
\cI_2 =  G^2 \sqrt{-g}  \frac{ \lrbrk{ g^{ij}g_{ij} - 2 } }{\det(g_{mn})}
	 + F^i_a F^j_b  \frac{ \sqrt{-g}g_{ij} g^{ab} }{\det(g_{mn})}
	 + 2F^j_a G  \frac{ \sqrt{-g}g_{ji} g^{a i} }{\det(g_{mn})},
\ee
where
\be
H_{ijk} \equiv \e_{ijk} G, \qquad H_{a ij} \equiv \e_{ijk} F^k_a,
\ee
and $g^{-1}_{cd}$ is the inverse of the $3\times 3$ matrix $g^{ab}$.
The action is invariant under the standard diffeomorphism transformations
associated with $x^k \rar x^k + \xi^k$, as well as under the modified diffeomorphism
\be \label{33diff}
\d B_{a i} = -\ove{2} \frac{\pa \cI_1}{\pa F^i_c} \e_{c b a} \xi^b
	-\ove{2} \frac{\pa \cI_2}{\pa F^i_c}\e_{cba} \xi^b, \quad
\d B_{ij} = \xi^b H_{bij},
\ee
associated with $x^b \rar x^b + \xi^b$.
For the diffeomorphism $x^b \rar x^b + \xi^b$,
$\d B_{ij}$ has the conventional form but $\d B_{a i}$ is modified and
reduces to the conventional form on the mass shell.

As we have already mentioned, the both actions \eqref{15diffS} and \eqref{33diffS},
and the modified diffeomorphisms \eqref{15diff} and \eqref{33diff}
can be obtained by an appropriate gauge fixing the corresponding covariant actions
\cite{Pasti:1996vs,Pasti:2009xc}.
Moreover, a generic crucial ingredient for the actions \eqref{15diffS} and \eqref{33diffS}
to enjoy the 6d diffeomorphism invariance is that the modified transformation laws be proportional to the derivatives of the corresponding actions
with respect to the gauge field strengths.
In our case of formally covariant 2+4 theory considered in Section \ref{sec:cov},
one can check that this property is lost in the ``gauge''
$a^s = \d^s_a x^a$
for the transformation rule \eqref{LPST24}.
Despite the mentioned difficulties in obtaining the modified diffeomorphism,
it is possible to show that coupling to gravity can be done,
which indirectly implies that there exists modified diffeomorphism for the
2+4 split model on a curved background. 
This encouragement comes from the Hamiltonian analysis which we will present right away. 

\subsection{A possible way out: Hamiltonian analysis}
The Hamiltonian analysis provides a natural way to put an action in curved space-time,
and to compute the (modified) diffeomorphism symmetry.
So it could provide a better insight into the issue with coupling of the $2+4$ model to 6d gravity.
This approach is adopted in the Henneaux-Teitelboim (HT) $1+5$ action \cite{Henneaux:1988gg}. 
We leave the systematic Hamiltonian analysis, following the work of \cite{Hojman:1976vp}, and the refinement of PST covariantisation 
of the 2+4 model as future works. 
Nevertheless, we will show that the ``gauge-fixed''\footnote{Readers should bear in mind that the (naive) PST covariantisation presented in section \ref{sec:cov} is not complete. 
However, one obtains a non-covariant action by naively setting $a^s = \d^s_a x^a$.} formally covariant action \eqref{LcovS} 
has the correct number of degrees of freedom by doing the Hamiltonian analysis. 
Moreover, the Hamiltonian density and momentum densities 
satisfy the hyper-surface deformation algebra, which suggests that 
the coupling to $6d$ gravity is promising.

Suppose we put $a^s = x^a\d_a^s$, $s=1,2$ (in the following $a,b,c,d=0,5$ and $i,j,k,l,m,n=1,2,3,4$) in the action \eqref{LcovS}, we obtain a non-covariant 
action with the Lagrangian density 
\be \label{gf24f}
\begin{split}
\cL &= -\ove{6}\e^{ab}\e^{klmn}H_{lmn}H_{ab k} + 2\e^{ab}G^i H_{b ij}g^{-1}_{ac}g^{c j}\\
    &\qquad -\frac{\sqrt{-g}}{2\det(g_{mn})} F_a^{kl}F_{b}^{ij}g^{ab}g_{ik}g_{lj} -2\frac{\sqrt{-g}}{\det(g_{mn})}G^l F_a^{k j}g^{a i}g_{k i}g_{l j}\\
    &\qquad -2\frac{\sqrt{-g}}{\det(g_{mn})}G^i G^j g^{kl}g_{l[k}g_{i]j} +\frac{\sqrt{-g}}{\det(g_{mn})}G^i G^j g_{ij}, 
\end{split}
\ee
where $G^i$ is defined as in \eqref{Gdef}, $g^{-1}_{ab}$ is the inverse of the $2\times2$ matrix 
$g^{ab}$, and 
\be
H_{0ij} = \ove{2}\e_{ijkl}F^{kl}_0, \qquad H_{5ij} = \ove{2}\e_{ijkl}F^{kl}_5. 
\ee
The conjugate momenta are 
\be
\begin{split}
\p^{ij} = \frac{\d\cL}{\d\dot B_{ij}}
= -2G^{[i}g^{-1}_{5c} g^{j]c} - \frac{\sqrt{-g}}{\det(g_{mn})} \lrbrk{\hlf\e^{klij}F_b^{pq}g^{0 b}g_{pk}g_{lq} +  G^l\e^{kqij}g^{0p}g_{kp}g_{lq}}
\end{split}
\ee
\be
\p^{5i} = \frac{\d\cL}{\d\dot B_{5i}} = -\Ht^{05i} = G^i, 
\ee
\be
\p^{0i} = \p^{05} = 0.
\ee
To go on with the analysis of 2+4 model coupling to gravity, 
we decompose the 6d metric according to Arnowitt-Deser-Misner-like Hamiltonian formalism
\be\label{metric}
g_{\m\n} =
\begin{pmatrix}
-(N^0)^2+\g_{\ah\bh}N^\ah N^\bh & \g_{\bh\ch}N^\ch\\
&\\
\g_{\ah\ch}N^\ch & \g_{\ah\bh}
\end{pmatrix}
.
\ee
We define the inverse of $\g_{\ah\bh}$ and its determinant as $\g^{\ah\bh},$ and $\g,$ respectively.
The inverse of the metric is
\be
g^{\m\n} =
\begin{pmatrix}
-(N^0)^{-2} & \frac{N^\bh}{(N^0)^2}\\
\frac{N^\ah}{(N^0)^2} & \g^{\ah\bh}-\frac{N^\ah N^\bh}{(N^0)^2}
\end{pmatrix}
.
\ee
The determinant is
\be
g = -(N^0)^2 \g
\ee
so
\be
\sqrt{-g} = N^0\sqrt{\g}.
\ee

After a somewhat lengthy calculation, the canonical Hamiltonian 
can be found to be 
\be
\begin{split}
\cH &= N^0\lrbrk{\hlf\frac{1}{\sqrt{\g}}\p^{ij}\p^{\ah\bh}\g_{\ah i}\g_{\bh j}
+\ove{\sqrt{\g}}\p^{j5}\p^{\ah\bh}\g_{\ah j}\g_{\bh 5}
+\hlf\ove{\sqrt{\g}}\Ht^{0\ah\bh}\Ht^{0\mh\nh}\g_{\ah\mh}\g_{\bh\nh}}\\
&\qquad
+\hlf N^\ih\e_{\ih\mh\nh\xh\yh}\Ht^{0\mh\nh}\p^{\xh\yh}
-\p^{ij}\pa_i B_{j0} - 2\p^{5i}(\pa_5 B_{i0} + \pa_i B_{05}), 
\end{split}
\ee
where 
the hatted Roman indices are 5d indices, $\ah,\bh,\mh,\nh=1,2,3,4,5$.
Note that this Hamiltonian is at most linear in $N^0, N^\ah.$
This form has the potential to not spoil the degrees of freedom counting for the gravity sector.
But to make sure, we have to check that the hyper-surface deformation algebra is really satisfied.
Let us denote
\be
\cH_0 = \hlf\frac{1}{\sqrt{\g}}\p^{ij}\p^{\ah\bh}\g_{\ah i}\g_{\bh j}
+\ove{\sqrt{\g}}\p^{j5}\p^{\ah\bh}\g_{\ah j}\g_{\bh 5}
+\hlf\ove{\sqrt{\g}}\Ht^{0\ah\bh}\Ht^{0\mh\nh}\g_{\ah\mh}\g_{\bh\nh},
\ee
\be
\cH_\ih = \hlf \e_{\ih\mh\nh\xh\yh}\Ht^{0\mh\nh}\p^{\xh\yh}
\ee

To couple the theory to 6d gravity, we consider the full Hamiltonian  
\be
\cH^{\text{full}} = \cH^{(g)} + \cH,
\ee
where the pure gravity Hamiltonian is given by 
\be
\cH^{(g)} = N^\m \cH_\m^{(g)},
\ee
with
\be
\cH_0^{(g)} = -\sqrt{\g}R + \ove{\sqrt{\g}}\lrbrk{\z^{\ah\bh}\z^{\mh\nh}\g_{\ah\mh}\g_{\bh\nh} - \hlf(\z^{\ah\bh}\g_{\ah\bh})^2},
\ee
\be
\cH_\ah^{(g)} = -2\g_{\ah\bh}\nabla_\ch\z^{\bh\ch}.
\ee
In the above expressions,
$\g_{\ah\bh}$ is (spatial) 5d metric,
$\z^{\ah\bh}$ is conjugate momentum to $\g_{\ah\bh},$
$R$ is 5d Ricci scalar,
and $\nabla_\ah$ is $\g-$compatible covariant derivative.

The primary constraints are 
\be
\P^{(g)}_\m \approx 0,\qquad
\p^{0\ah} \approx 0,\qquad
\p^{5i}  + \Ht^{05i} \approx 0,
\ee
where $\P^{(g)}_\m$ is conjugate to $N^\m$ and $\approx$ denotes a weak equality 
which only holds on the constraint surface. 
The secondary constraints include 
\be
{\p}^{ij}  + \Ht^{0ij} \approx 0, \qquad \pa_{\mh} {\p}^{\mh\ah} \approx 0, \qquad 
\cH_\m^{(g)} + \cH_\m \approx 0. 
\ee
Among which, we have first-class constraints 
\be
\P_\m\approx 0\ (6),\qquad
\p^{0\ah}\approx 0 \ (5),\qquad
(\cH_\m^{(g)} + \cH_\m)\approx 0 \ (6),\qquad
\pa_{\mh} \p^{\mh\ah}\approx 0 \ (4), 
\ee
where the numbers in the parenthesis indicate the number of 
the corresponding independent constraints. 
On the other hand, we also have 6 second-class constraints which is the transverse components of 
${\pi}^{\mh\nh} + \Ht^{0\mh\nh} \approx 0$.

Let us count the number of degrees of freedom.
There are $72$ phase space variables, $42$ of them coming from gravity sector
while $30$ of them coming from gauge sector.
There are $21$ first-class constraints and $6$ second-class constraints.
Therefore, the number of degrees of freedom is given by
\be
\begin{split}
\text{number of degrees of freedom} 
&=\frac{(42+30) - 2\times 21 - 6}{2}\\
&=12\\
&=9+3. 
\end{split}
\ee
Note that graviton in 6d has $9$ degrees of freedom, therefore 
the calculation shows that the 2-form theory \eqref{gf24f} indeed has the desired 3 
degrees of freedom. 

When classifying class of the constraint, we have considered
Poisson's brackets between the constraints.
Let us list only the hyper-surface deformation algebra:
\bea
[\cH^{\text{full}}_0(x),\cH^{\text{full}}_{0}(x')] &=& (\g^{\ah\bh}(x)\cH^{\text{full}}_\ah(x)+\g^{\ah\bh}(x')\cH^{\text{full}}_\ah(x'))\pa_\bh\d^{(5)}(x,x'),\\\
[\cH^{\text{full}}_\ah(x),\cH^{\text{full}}_0(x')] &=& \cH^{\text{full}}_0(x)\pa_\ah\d^{(5)}(x,x')\nn\\
&&\quad+\frac{2}{\sqrt{\g(x)}}\pa_{\mh}{\p}^{\mh\nh}(x){\p}^{\bh\ch}(x)\g_{\bh\ah}(x)\g_{\nh\ch}(x)\d^{(5)}(x,x'),\\\ 
[\cH_\ah^{\text{full}}(x),\cH^{\text{full}}_\bh(x')] &=& \cH^{\text{full}}_\ah(x')\pa_\bh\d^{(5)}(x,x')+\cH^{\text{full}}_\bh(x)\pa_\ah\d^{(5)}(x,x')\nn\\
&&\quad+\pa_{\mh}\p^{\mh\nh}(x)\e_{\bh\jh\kh\ah\nh}(x)\Ht^{0\jh\kh}(x)\d^{(5)}(x,x'),
\eea
where $\cH^{\text{full}}_\m = \cH^{(g)}_\m + \cH_\m.$
Note that time dependence in the above formula
is suppressed since Poisson bracket is computed at equal time
and that $x$ and $x'$ represent 5d spatial coordinates.
We see that the above Poisson's brackets weakly vanish,
indicating that the hyper-surface deformation algebra is satisfied.

The correctly obtained number of degrees of freedom
and hyper-surface deformation algebra tell us
that the couple of the quadratic $2+4$ action to gravity is 
actually doable. 
The full systematic Hamiltonian analysis and the refinement of PST covariantisation 
on 2+4 are left in the upcoming works.

\setcounter{equation}0
\section{Conclusion}\label{sec:conc}

In this paper we have analysed a possibility of supersymmetrising and coupling to gravity the free theory for the 2--form chiral gauge field in six--dimensional space--time in the formulation with the manifest $SO(1,1)\times SO(4)$ invariance \cite{Chen:2010jgb} and generalize it to include non--linear self--interactions of a Born--Infeld type.
In the formulation with the 2+4 split space--time we have {\color{black}constructed} an action describing $\cN = (2,0)$ tensor supermultiplet.
On the other route, we have constructed a non-linear Lagrangian for the chiral 2-form in $d=2+4$ with a non-manifest 6d Lorentz invariance, whose equations of motion amount to the non-linear self-duality condition
which coincides with that obtained from the superembedding description of the dynamics of the M5--brane.

In order to make a further extension of these results and ultimately obtain the complete M5-brane action in 6d space--time with 2+4 splitting, one should couple the 2+4 action to 6d gravity, using e.g. the PST technique. However, our analysis showed that
the covariantisation of this system via the conventional PST approach
does not work, at least straightforwardly. 
Nevertheless, the non-covariant theory obtained by naively gauge-fixing 
has the correct number of degrees of freedom. 
Though the straightforward application of PST technique is not successful, 
the counting of the number of degrees of freedom suggests that we are on the right track.

Having encountered the above mentioned issues in one of the alternative Lagrangian formulations for the 6d chiral gauge field, it would be of interest to study if similar difficulties arise in self-dual Lagrangian descriptions of chiral gauge fields with different splittings of space--time in other dimensions \cite{Chen:2010jgb,Huang:2011np}.

\subsection*{Acknowledgements}
We are very grateful to Dmitri Sorokin for the collaboration at the initial stage of the work,
various helpful comments, and comments on the manuscript.

The authors are also grateful to Igor Bandos, Chong-Sun Chu,
 Pei-Ming Ho, Hiroshi Isono, Gaurav Narain, Paolo Pasti, Douglas Smith, Shingo Takeuchi, Tomohisa Takimi and Mario Tonin for discussions.

Sh-L. K. is grateful to Chulalongkorn University for kind hospitality extended to him
during the 4th Bangkok Workshop on High-Energy Theory while working in progress
and kind hospitality of National Center for Theoretical Sciences (NCTS) Physics Division (Taiwan) where part of the work was presented.
P.V. acknowledges the support from Naresuan University grant R2558C144.
P.V. would like to acknowledge hospitality and support extended to him by Durham University
during an initial stage of this project,
the 4th Bangkok Workshop on High-Energy Theory at Chulalongkorn University (Bangkok, Thailand, 19th-27th January 2015)
and NCTS (Hsinchu, Taiwan, 5th May-14th June 2015)
during work in progress.

\appendix

\setcounter{equation}0
\section{Equivalence of self-duality equations}\label{app:SDSE}

To show that the self-duality equations derived from the action
\eqref{S24}
are equivalent to the ones in the super-embedding approach \eqref{embsd},
we should check that
\bea
 H^-_{a ij}
 &=& 4Q^{-1}\lrsbrk{4g^2 h_{a ij} + 8g^mg_ih_{a jm} - 8g^m g_j h_{a im} + 2h_{a xy}h^{b xy}h_{b ij}} \nn\\
 &=& -\ove{4}\frac{\pa\cI}{\pa H^{+a ij}} \lrbrk{ H^+(h,g), G(h,g) },
\eea
\be
\Gt^l = 4 Q^{-1}\lrsbrk{(1-4g^2)g^l + 4 g^x h_{a xk}h^{a lk} }
= \ove{2} \frac{\pa\cI}{\pa G_l} \lrbrk{ H^+(h,g), G(h,g) }.
\ee
That is, we need to calculate the derivatives of $\cI$ with respect to $H^+$ and $G$,
and then express $H^+$ and $G$  in terms of $h$ and $g$, as in \eqref{SEH+} and \eqref{SEG},
to check that the results coincide with \eqref{SEH-} and \eqref{SEGt}.

When dealing with the self-dual tensor $H^+_{a ij}$, we found it convenient
to further split its components into independent ones and utilize a `bra-ket' notation
\be
F^{\pm ij} \equiv H^{\pm5ij}, \quad G_k \rar \ket{G}_k, \quad h^{5ij} \equiv f^{ij}, \quad g_k \rar \ket{g}_k.
\ee
$f^n$ implies matrix multiplication, e.g. $f^{ij}f^{jk}$ etc., $g^2\equiv g_i  g^i=\langle{g}|{g}\rangle$, and $f\ket{g}$ and $\bra{g} f$ stand, respectively, for
$f^{ij}g_j$ and $g_j f^{ji}$.

In this notation, the super-embedding duality equations take the form
\be \label{Finf}
F^+ = 4Q^{-1}f,
\ee
\be \label{Fmmatrix}
F^- = 4Q^{-1} \Big( (2\tr f^2+4g^2)  f - 8f^3 - 8\ket{g}\bra{g}f - 8 f\ket{g}\bra{g} \Big),
\ee
\be  \label{Gket}
\ket{G} = 4Q^{-1} \Big( (1+4g^2-2\tr f^2) \ket{g} + 8f^2\ket{g} \Big),
\ee
\be \label{Gtket}
\ket{\Gt} = 4Q^{-1} \Big( (1-4g^2+2\tr f^2) \ket{g} - 8f^2\ket{g} \Big),
\ee
\be
Q = 1 - 16(g^2)^2 + 16y + x.
\ee
The $X,Y$ and $G^2$, eq. \eqref{X}, and $x,y$, eq. \eqref{xy}, can be written as
\be
x = 4\lrbrk{ \tr f^2}^2 - 16 \tr f^4, \quad y = g^2\tr f^2 - 4 \bra{g}f^2\ket{g},
\ee
\be
X = 4\lrbrk{ \tr F^{+2}}^2 - 16 \tr F^{+4}, \quad Y = G^2\tr F^{+2} - 4 \bra{G}F^{+2}\ket{G}.
\ee
The self-duality equations derived from the action principle take the form
\be \label{SDbraket}
\ket{\Gt} = \ove{2}\ket{ \frac{\pa\cI}{\pa G} }, \qquad
F^- = -\ove{4}\frac{\pa\cI}{\pa F^+}.
\ee
One can calculate the derivatives of $\cI$ in a straightforward though tedious way.
In particular, one needs to obtain
$$
\frac{\pa Q}{\pa G^2}, \quad \frac{\pa Q}{\pa X}, \quad \frac{\pa Q}{\pa Y},
$$
from the sextic equation \eqref{sextic}.
The argument of the square root $\sqrt{1+G^2-Q^2X/64}$ becomes a perfect square
when the expressions
\eqref{xX} and \eqref{G2se} are used
\be
\sqrt{1+G^2-Q^2X/64}
= \sqrt{\frac{(-8 {g^2}+Q-2)^2}{Q^2}}
= \frac{8g^2 +2-Q}{Q}.
\ee
Then, the checking of \eqref{Gtket} is straightforward,
while the checking of \eqref{Fmmatrix} is a bit more complicated.
After substituting \eqref{Finf} and
\eqref{Gket} into
the derivative $\pa\cI/\pa F^+$,
we see that we need to deal with terms of the following form
\be \label{ugbasis}
f^2\ket{g}\bra{g}f  \quad\text{and}\quad f^2\ket{g}\bra{g}f^3.
\ee
These terms can be traded with other simpler-looking terms as follows.
Applying to $\bra{g}f^4\ket{g}$ and $\bra{g}f^6\ket{g}$ the Cayley-Hamilton formula
\be
M^{4} = \hlf \lrbrk{ \tr M^{2} } M^{2} - \lrbrk{-\ove{4}\tr M^{4}+\ove{8}\lrbrk{\tr M^{2}}^2},
\ee
where $M$ is any  anti-symmetric $4\times 4$ matrix,
we have
\be
\bra{g}f^4\ket{g} = \ove{2}\tr f^2 \, \bra{g}f^2\ket{g} - \lrbrk{ \ove{8}(\tr f^2)^2 - \ove{4}\tr f^4 } g^2.
\ee
Taking derivatives of the both sides of the above equations,
we can then trade $f^2\ket{g}\bra{g}f$
with another more convenient basis.
For example,
\be
\begin{split}
&\ket{g}\bra{g}f^{3}+f^{3}\ket{g}\bra{g}+f^{}\ket{g}\bra{g}f^{2}+f^{2}\ket{g}\bra{g}f^{}\\
&= f\bra{g}f^{2}\ket{g}+\hlf\tr f^{2}\lrbrk{ \ket{g}\bra{g}f + f\ket{g}\bra{g} }
+\hlf(2 f^{3}-\tr f^{2}f)\ipr{g}{g}.
\end{split}
\ee
This identity also implies that the terms on the left-hand-side of the above equation
always show up together.
Terms like $f^2\ket{g}\bra{g}f^3$ can be simplified by utilising the above
identity and then using Cayley-Hamilton theorem repeatedly if necessary.

In this way, we have checked, using Mathematica, that
\eqref{Fmmatrix} and \eqref{Gtket} are correctly reproduced by \eqref{SDbraket}.

\setcounter{equation}0
\section{Checking the worldvolume space-time symmetry constraint}\label{app:SC}
{\color{black}
 {\color{black}Without loss of generality,}
the {\color{black}worldvolume} space-time symmetry constraint \eq{ncSC} on the form of the nonlinear self-dual action reduces to
\be
2F^{+ij}G_j - F^{+ij}\frac{\pa \cI}{\pa G^j} - \frac{1}{2}\lrbrk{\frac{\pa \cI}{\pa F^+}}^{ij}G_j
-\ove{4} \lrbrk{\frac{\pa \cI}{\pa F^+}}^{ij}\frac{\pa \cI}{\pa G^j} = 0,
\ee
where $F^{+ij}\equiv H^{+5ij}$.

In order to proceed, it is convenient to utilize the `bra-ket' notation,
introduced in the appendix \ref{app:SDSE}.
Then, the above equation can be written as follows
\be\label{consFGbk}
2F^+\ket{G}-F^{+}\ket{\frac{\pa \cI}{\pa G}} - \frac{1}{2}\frac{\pa \cI}{\pa F^+}\ket{G}
-\ove{4}\frac{\pa \cI}{\pa F^+}\ket{\frac{\pa \cI}{\pa G}} = 0,
\ee
where the derivatives of $\cI$ with respect to $G_i$ and $F^+_{ij}$ have the form
\be
\frac{\pa\cI}{\pa G_i}
= \frac{\pa G^2}{\pa G_i}\frac{\pa\cI}{\pa G^2}
   + \lrbrk{ \frac{\pa G^2}{\pa G_i} \frac{\pa Q}{\pa G^2} + \frac{\pa Y}{\pa G_i} \frac{\pa Q}{\pa Y} } \frac{\pa\cI}{\pa Q},
\ee
\be
\frac{\pa\cI}{\pa F^+_{ij}}
= \frac{\pa X}{\pa F^+_{ij}}\frac{\pa\cI}{\pa X}
   + \lrbrk{ \frac{\pa X}{\pa F^+_{ij}} \frac{\pa Q}{\pa X} + \frac{\pa Y}{\pa F^+_{ij}} \frac{\pa Q}{\pa Y} } \frac{\pa\cI}{\pa Q},
\ee
and the derivatives of $Q$ with respect to $X,Y$ and $G^2$ can be obtained from the sextic equation \eqref{sextic}.

The left hand side of \eqref{consFGbk}
{\color{black}can} be expressed in terms of the two-vector basis:
\be
F^3\ket{G}, \qquad F\ket{G},
\ee
with complicated coefficients{\color{black},}
{\color{black}
which are fractions and contain $\sqrt{64+64G^2 - Q^2 X}.$
In order to proceed, we make a common denominator for the both coefficients,
and call $C_1$ the numerator of the coefficient of $F\ket{G}$,
and $C_3$ the numerator of the coefficient of $F^3\ket{G}$.
}

To show that \eqref{consFGbk} is satisfied, we should check that $C_1=C_3=0.$}
Assuming that $C_1$ vanishes,
we can obtain the expression for the square root by solving the corresponding equation
\be
C_1(\sqrt{\cdots}, Q,X,Y,G^2,\tr F^2) = 0 \quad \Leftrightarrow
\sqrt{64+64G^2 - Q^2 X} = D_1(Q,X,Y,G^2,\tr F^2),
\ee
where $D_1$ is a fraction composed of $Q,X,Y,G^2$ and $\tr F^2$.
This requirement can then be easily rearranged into a $Q^n$ series equation of the form
\be\label{candidv2}
\sum_{n=0}R_n(X,Y,G^2,\tr F^2)Q^n = 0.
\ee
This candidate identity will be \emph{trivially} satisfied if and only if $C_1$ is zero.
We then simplify the candidate identity by trading all the $Q^n$ with $n\geqslant6$
in terms of the sextic equation \eqref{sextic}, with lower degrees of $Q$.
The final result is that \eqref{candidv2} is indeed the identity.
The check that $C_3=0$ is carried out in a similar way.
Therefore, the nonlinear self-dual action \eqref{S24} with $Q$ satisfying the sextic equation \eqref{sextic}
indeed has the (modified) {\color{black}worldvolume} space-time symmetry.

\begin{comment}
\bibliographystyle{utphys}
\bibliography{references1}
\end{document}